\documentclass[11pt,letterpaper,superscriptaddress,nofootinbib]{revtex4}
\usepackage{amsfonts,amsmath,amsthm,amssymb,amscd}
\usepackage{color,graphicx,array,tikz,pgfplots}
\usepackage[scaled]{helvet} % ss
\usepackage{courier} % tt
\usepackage{eulervm} % a better implementation of the euler package (not in gwTeX)
\normalfont
\usepackage[T1]{fontenc}

\newcommand{\eq}[1]{(\ref{#1})}
\newcommand{\sect}[1]{\S\ref{#1}}

\begin{document}

\title{Fast Optimization Algorithms and the Cosmological Constant}
\date{June 23, 2017}

\author{Ning~Bao}
\affiliation{\small{Institute for Quantum Information and Matter and Walter Burke Institute for Theoretical Physics, California Institute of Technology, Pasadena, CA 91125}}
\author{Raphael~Bousso}
\affiliation{\small{Center for Theoretical Physics and Department of Physics, University of California, Berkeley, CA 94720}}
\affiliation{\small{Lawrence Berkeley National Laboratory, Berkeley, CA 94720}}
\author{Stephen~Jordan}
\affiliation{\small{Joint Center for Quantum Information and Computer Science, University of Maryland, College Park, MD 20742}}
\affiliation{\small{National Institute of Standards and Technology, Gaithersburg, MD, 20899}}
\author{Brad~Lackey}
\affiliation{\small{Joint Center for Quantum Information and Computer Science, University of Maryland, College Park, MD 20742}}
\affiliation{\small{Departments of Computer Science and Mathematics, University of Maryland, College Park, MD 20742}}
\affiliation{\small{Mathematics Research Group, National Security Agency, Ft.~G.~G.~Meade, MD 20755}}

\begin{abstract}
Denef and Douglas have observed that in certain landscape models the problem of finding small values of the cosmological constant  is a large instance of an NP-hard problem. The number of elementary operations (quantum gates) needed to solve this problem by brute force search exceeds the estimated computational capacity of the observable universe. Here we describe a way out of this puzzling circumstance: despite being NP-hard, the problem of finding a small cosmological constant can be attacked by more sophisticated algorithms whose performance vastly exceeds brute force search. In fact, in some parameter regimes the average-case complexity is polynomial. We demonstrate this by explicitly finding a cosmological constant of order $10^{-120}$ in a randomly generated $10^9$-dimensional ADK landscape.
\end{abstract}

\maketitle

%\tableofcontents

\section{Introduction and Summary}

\subsection{Cosmological Constant Problem and the Landscape}

According to the Standard Model of particle physics, the energy density of the vacuum receives multiple contributions whose order of magnitude vastly exceeds the observed value~\cite{Per98,Rie98,Planck}
\begin{equation}
\Lambda \approx 1.5\times 10^{-123} M_P^4~.
\end{equation}
(Below we will use units where the Planck mass is unity, $1\equiv M_P=(\hbar c/G)^{1/2}\approx 1.2 \times 10^{19}$ GeV.) Both perturbative and nonperturbative processes contribute, such as vacuum fluctuations of all fields, and electroweak symmetry breaking.  The excess is by a factor of at least $10^{60}$ assuming a new symmetry at a TeV (so far not found). It could be as large as $10^{122}$ with a Planck-scale cutoff. The observed small value of $\Lambda$ implies that the various contributions must cancel against one another, or against further unknown contributions which must be at least as large, with a relative precision of at least $10^{-60}$ and perhaps $10^{-122}$. 

Consistency with well-established cosmological history severely constrains large classes of approaches to this problem. For example, it is not possible for the universe to dynamically select the ``correct'' vacuum energy at early times. Only gravity couples to the absolute energy, and gravity sees the total stress tensor. At the time of Big Bang Nucleosynthesis, characteristic energy densities were of order $10^{-88}$. This is more than 30 orders of magnitude greater than the observed value that would have to be targeted by a putative adjustment mechanism. Attempts to desensitize General Relativity to the energy in vacuum fluctuations run into conflict with tests of the equivalence principle. These and other obstructions to non-anthropic approaches are discussed in~\cite{Pol06,Bou07}.

In a landscape model, a small cosmological constant is selected by correlation with the location of observers. The universe can form large regions with many different possible values of $\Lambda$. This is most natural in a theory with extra dimensions, such as string theory. One finds that there are generically exponentially many ways of constructing a ``vacuum'', i.e., a compactification to 3 large spatial dimensions. If the vacuum energy $\Lambda$ is, say, a random number between $-1$ and $1$, but there are ${\cal N} \gg 10^{122}$ different vacua, it is likely that a small fraction but large number $10^{-122} {\cal N}$ of vacua have small enough $\Lambda$ to be consistent with observation. Moreover, a great variety of vacua are naturally produced by inflationary dynamics in the early universe. In specific models, the distribution of $\Lambda$ is not random. The above approach works as long as the spectrum of $\Lambda$ is sufficiently dense near $0$. Consistency with standard cosmological history is achieved if the potential landscape is multi-dimensional, with neighboring vacua generically having very different energies~\cite{BP}.

Typical spacetime regions would still have $\Lambda\sim O(1)$, of course. But in such regions any worldline has an event horizon of order the Planck area, and so contain only a few bits of causally connected information~\cite{GibHaw77a,CEB1}. Complex structures such as observers necessarily find themselves in a highly atypical region that allows for a larger cosmological horizon with area (and hence, maximum entropy) of order $\Lambda^{-1}$. (The origin of the particular scale $10^{-122}$ is not explained by this qualitative argument. See \cite{Wei87} for an argument that assumes galaxies are needed, or \cite{BouFre10d} for a more robust argument.)

\subsection{Computational Complexity}

In 2007, Denef and Douglas brought a complexity theoretic perspective to the cosmological constant problem \cite{DD}. In particular, they pointed out that, in some formulations, the problem of finding a vacuum with cosmological constant compatible with observation is a large instance of an NP-hard problem. Specifically, two simplified models were considered in \cite{DD}: a version of the Arkani-Hamed-Dimopolous-Kachru (ADK) model \cite{ADK}, and the Bousso-Polchinski (BP) model \cite{BP}. Here we focus on the ADK model, which is the more simplified of the two, as it is sufficient to capture the essential features that we wish to address.

In the ADK model, the cosmological constant is obtained by summing the energy contributions from a large number of fields, each of which is subject to a double-well potential. We assume the vacuum energy contributed by either of the two minima of each field to be a random number with mean zero\footnote{This assumption differs from the model mainly studied by ADK, but it is adequate for our analysis.} and standard deviation of of order 1 in Planck units. (Thus it can be positive or negative.) Given $n$ such fields there are correspondingly ${\cal N}=2^n$ metastable vacua, specified by an $n$-bit string $f(j) \in \{0,1\}$, $j=1,\ldots, n$. The cosmological constant in any vacuum is given by
\begin{equation}
\label{def}
    \Lambda[f(j)] = \sum_{j=1}^n E_{f(j)}^{(j)}
\end{equation}
where $E_0^{(j)}$ and $E_1^{(j)}$ are the two possible vacuum energies contributed by the $j^{\mathrm{th}}$ field.

If our universe were described by this model, then with appropriate technology, there would be no obstruction in principle to measuring each of the $n$ fields directly, and thus determining which of its two vacua it occupies. This requires only $n$ measurements. Thus, we can in principle identify which vacuum we live in, among all the vacua in the ADK model. A similar argument applies to the BP model: given good enough technology, one would simply measure the fluxes on topological cycles in the extra dimensions. We could probe each field experimentally and read off the bit string $f(j)$. 

Denef and Douglas consider a different task: suppose we are given only the total value of the cosmological constant $\sim 10^{-122}$ (for example from observation), but not the vacuum configuration $f(j)$ of the $n$ fields. We wish to identify a vacuum in the ADK model compatible with this value. Then we would have to sift through the $2^n$ allowed vacua to find a combination of positive and negative numbers, each of order 1, that add up to $10^{-122}$. Such combinations clearly constitute a small fraction of all the $2^n$ vacua. However, in simple statistical models, \emph{e.g.}\ where $E_0^{(1)},E_1^{(1)},\ldots,E_0^{(n)},E_1^{(n)}$ are each independently drawn uniformly at random from $[-1,1]$, such combinations will exist with high probability provided $\sqrt{n}\, 2^{-n} \lesssim 10^{-122}$ \cite{KKLO86}, \emph{i.e.}\ $n \gtrsim 407$. Furthermore, for $n$ larger than this, the number of vacua with $\Lambda \leq 10^{-122}$ will be roughly $10^{-122} \times 2^n/\sqrt{n}$~\cite{mertens2000random, bauke2004number}.

In \cite{DD} it was pointed out that the problem of finding such vacua in the ADK model is a variant of the number partitioning problem, which is NP-complete. Consequently, under the widely-held complexity-theoretic assumption that $P \neq NP$, no classical algorithm can solve worst-case instances of this problem in time scaling polynomially with $n$. Furthermore, under the stronger but also widely-held assumption that $NP \nsubseteq BQP$, no quantum algorithm can solve worst-case instances of this problem in polynomial time either. 

The physical significance of the Denef-Douglas observation is not immediately clear. 
Here, we posit that its significance lies in the contrast between the NP-complete hardness of finding a vacuum with small $\Lambda$ by studying the theory, on the one hand; and on the other hand, the ease with with we can read off a solution to this problem (our own vacuum), by measuring the $n$ bits directly as discussed above. This implies that we get to read off the answer to an instance of an NP-hard problem that Nature has already solved for us. And we get to do this for anthropic reasons: complex structures exist only in regions with $\Lambda\ll 1$. Our mere status as observers gives us immediate access to the solution of a hard problem. How is this possible? 

It is instructive to consider the cosmological dynamics that had to solve the ``hard'' problem and produce the small-$\Lambda$ region we occupy. There are two valid and largely equivalent~\cite{Bou09} viewpoints, global and local. In the global viewpoint, the universe is exponentially expanding and constantly producing new regions. In this case gravity supplies exponential resources for solving the hard problem. No-one can observe the whole universe, because regions are shielded from one another by event horizons. But observers necessarily find themselves in the regions where the problem has been solved.

In the local viewpoint, one considers the different decay chains through the landscape that might be realized in a {\em single} causally connected region (causal patch). The patch decoheres rapidly every time a vacuum transition takes place. This trades the multiverse for ``many worlds''~\cite{BouSus11}. Observers find themselves in a branch of the decay chain that produced a vacuum with small $\Lambda$. The situation is comparable to solving a hard problem by sitting down in front of a robot that points a gun at you. The robot takes one random guess (generated by some quantum measurement) and secretly checks it in polynomial time. If the guess solves the problem, the robot tells you the solution, but if it fails, it shoots you. Necessarily, if you survive, you will have gained the solution very quickly\footnote{This method of solving NP-complete problems seems to have been first proposed in \cite{Moravec}; see also \cite{Aaronson05}.}.

We do not claim that from either of those viewpoints, our easy access to a solution of a hard problem constitutes a logical contradiction. Yet, the ability to utilize exponential unobservable resources, or an exponentially large branching tree of decoherent histories would be a surprising and perhaps troubling circumstance. Therefore, in this paper, we will posit a \emph{Computational Censorship Hypothesis}: by physical measurements we must not be accessing the solution to a hard problem, i.e., a problem so hard that it could not have been solved by the physical resources in the observable universe.

By ``resources,'' we mean the number of elementary gates in a computation. There is some ambiguity how to quantify an upper bound on this for the observable universe. Possible candidates include (in natural units) the Einstein-Hilbert-matter action~\cite{BroRob15}; the energy of the universe times its age~\cite{Lloyd}; the maximum entropy of the visible universe~\cite{CEB1,CEB2} or of any universe with the observed value of $\Lambda$~\cite{Bou00a} (which is given by the horizon area of empty de~Sitter space~\cite{GibHaw77a}); or lastly the amount of entropy that has been produced in our past light-cone. All but one of these definitions give a number of gates of order $\Lambda^{-1}\sim 10^{122}$ for our universe in the present era. (The final definition gives a somewhat lower answer~\cite{BouHar07} if event horizons are not included.) 
Thus, for the purposes of this paper, we will take the available resources to be:
\begin{equation}
R_{\rm max}\sim \Lambda^{-1}
\end{equation}
quantum gates. (Whereas this estimate takes an elementary quantum gate to be the notion of computational step relevant to our universe, other more speculative possibilities have been considered elsewhere \cite{AW09,BLSS09,BW15,AL98,Bao:2015hdp}.)

We note that making the Computational Censorship Hypothesis precise is a difficult problem that we don't claim to have solved. The central difficulty is that our universe provides us with the solution to one \emph{instance} of a hard problem, whereas computational complexity is defined only for asymptotic families of instances. For any instance of a problem there always exists an efficient algorithm which has the solution to that instance hardwired in. (We thank S.~Aaronson for stressing this point to us.) In an intuitive sense, it is clear that the existence of such algorithms is not of interest in determining the difficulty of the instance. Instead we take the complexity of the instance to be the number of steps required by the most efficient general-purpose algorithm that solves it. The distinction between general-purpose algorithms and ones with answers hard-wired seems difficult to formalize, but is typically easy to make in practice. 

In the remainder of this paper we will describe various general-purpose number partitioning algorithms that set upper bounds on the complexity of number partitioning problems. Different algorithms provide the best upper bound in different parameter regimes. In all regimes we find that the complexity of the cosmological constant problem within the ADK model is well within the computational capacity of the observable universe and therefore, contrary to initial appearances based on brute force search, it does not pose a challenge to the Computational Censorship Hypothesis. In some regimes the speedup over brute search achieved by more sophisticated algorithms is quite dramatic; for instances in which the ADK model has $10^9$ fields we are able to find a cosmological constant of order $10^{-120}$ in a few hours on a single processor.

Note that the Computational Censorship Hypothesis is quite minimal. We require only that some algorithm exists that can solve the problem (e.g., identify a suitable vacuum) in $10^{122}$ steps or less. We do not require that this algorithm bear any relation to the (largely known) cosmological dynamics that would have produced our universe. By contrast, recent work of Denef, Douglas, Greene, and Zukowski explores computational complexity as a possible restriction on the dynamics~\cite{DenDouTA,DenDouGreZuk17}. A related but distinct principle was proposed by Aaronson \cite{Aaronson05}, that NP-complete problems should not be solvable with polynomial resources by any physical means. Recent applications of this and related principles include \cite{Harlow:2013tf, Bao:2015hdp, Bao:2016uan}.

\subsection{An Apparent Paradox and Its Resolution}

Imposing the Computational Censorship Hypothesis leads to an apparent paradox in light of the Denef-Douglas result. To see this, we must quantify the hard problem and show that it requires resources larger than $R_{\rm max}\sim \Lambda^{-1}$. Indeed, as shown in section \ref{cost}, the number of elementary computational steps (quantum gates) required to find a solution with $\Lambda \sim 10^{-122}$ by brute force search of the landscape scales as 
\begin{equation}
R_{\rm brute}\sim \Lambda^{-1} \left( \log_2 \Lambda^{-1} \right)^{3/2}~, 
\end{equation}
which is asymptotically larger than the computational capacity $~\Lambda^{-1}$ in the limit of small $\Lambda$. For the particular value of $\Lambda \sim 10^{-122}$,  $\Lambda^{-1} \left( \log_2 \Lambda^{-1} \right)^{3/2}$ exceeds $\Lambda^{-1}$ by several orders of magnitude.

If the complexity of brute force search were the correct measure of the complexity of the number partitioning problem, then by measuring which vacuum we are in (which is in principle possible, as argued above) we would obtain the solution to an instance of a computational problem which could not be solved within our observable universe, in violation of the Computational Censorship Hypothesis. Furthermore, this violation does not necessarily require any measurements beyond present-day capabilities. The decision version of the number partitioning problem, of determining whether a solution with residue smaller than a give threshold exists, is already NP-hard, even without demanding that the explicit solution be produced. Thus, if we knew the specifics of the problem instance ($E_0^{(1)},E_1^{(1)},\ldots,E_0^{(n)},E_1^{(n)})$, then the astronomical observations that have already been made, indicating that $\Lambda \simeq 10^{-122}$ already tells us that a residue of that magnitude exists among the solutions to this instance of number partitioning, thereby learning the solution to a large instance of an NP-hard problem.

In the remainder of the paper we will examine how this apparent paradox can be resolved.
Our key observation is that modern algorithms can solve the number partitioning problem using far fewer computational steps than are required by brute-force search. The fastest known classical algorithm for general instances of the number partitioning problem runs in $R\sim O(2^{0.291n})$ time \cite{BCJ} and the fastest known quantum algorithm runs in $R\sim O(2^{0.241n})$ time \cite{BJLM}. For $n \lesssim 1300$ these algorithms place the instance of number partitioning arising in the ADK model within the estimated computational capacity of the observable universe, but far outside the capacity of even the largest supercomputers. 

Interestingly, for very large $n$, the problem becomes solvable with high probability by the Karmarkar-Karp heuristic, which runs in polynomial time,
\begin{equation}\label{a}
    R_{\rm KK}\sim n\log n~,
\end{equation}
provided that the number of numbers is sufficiently large,
\begin{equation}
    n \gtrsim \exp\left[\sqrt{\frac{\log B}{c}}\,\right]~,~~ c\approx 0.7~,
\end{equation}
where $B$ is the typical magnitude of the numbers. In the application to the ADK model,
\begin{equation}
B\sim \Lambda^{-1} \approx 10^{122}~.
\end{equation}
By exploiting the Karmarkar-Karp algorithm, we show in \sect{sec:experiments} that vacua with $\Lambda \sim 10^{-120}$ can in fact be found in the ADK model in under 3 hours on a standard workstation, provided 
\begin{equation}\label{b}
    n \gtrsim 10^9~.
\end{equation} %$n \gtrsim 7.5 \times 10^8$. 
While the worst-case remains NP-hard, Monte Carlo generated average cases can be solved in polynomial time, provided the number of fields is sufficiently large. 

In this work we have focused on the ADK model of the landscape which leads to number partitioning as the underlying computational problem. Karmarkar-Karp is a powerful algorithm against this problem, but it does not generalize to more complex models easily. It will be interesting to investigate the constraints imposed by the Computational Censorship Hypothesis on other toy models, such as the lattice model of BP which is not amenable to a Karmarkar-Karp style algorithm. Eventually one would hope to consider a concrete landscape arising from a complete theory, which would dictate both the structure of the partitioning problem and the statistical distribution of the input. For example the full string landscape~\cite{BP,KKLT}, when its structure becomes better understood, should provide data analogous to the concrete distribution of charges in the BP model.

Our results show that landscape models remain a viable approach to the cosmological constant problem even if the Computational Censorship Hypothesis is adopted. But for now, at least, we cannot confront the hypothesis specifically with the landscape of string theory, for three main reasons. First, the ADK model is purely a toy model; we know of no evidence that it arises from string theory. Second, the string landscape is understood only in a few corners of the theory, where small parameters are available and statistical estimates are arguably under control. In particular, the oft-quoted number $10^{500}$ of vacua is likely an underestimate~\cite{DouKac06}, and we do not know of a reliable upper bound. Third, even if we did know the structure of the landscape, and supposing that we knew of no general purpose algorithm that satisfied the Cosmic Censorship Hypothesis, this would not imply that no such algorithm exists. 

\paragraph*{Outline.} In section \ref{sec:number_adk} we relate the ADK model to number partitioning and estimate the brute force cost of finding a small value of $\Lambda$. In section \ref{sec:alg} we review the Karmarkar-Karp and other fast algorithms and discuss their range of applicability. In section \ref{sec:experiments} we report an empirical test of the Karmarkar-Karp algorithm. We demonstrate that it can find a value of $\Lambda$ consistent with observation in randomly generated instances of an ADK model with nearly $10^9$ fields (and so by Eq.~\ref{a}, in a few hours on a desktop computer). We find that sieves are less efficient but still suffice to demonstrate consistency with the Computational Censorship Hypothesis.

\section{Complexity of the ADK Model}
\label{sec:number_adk}

In this section, we show that the problem of finding a small cosmological constant $\Lambda$ in the ADK model can be reduced to the standard number partitioning problem. We then demonstrate that the cost of a brute force search exceeds $\Lambda^{-1}$ by a factor $(\log_2 \Lambda^{-1})^{3/2}$. Therefore a brute force search is incompatible with the Computational Censorship Hypothesis.

\subsection{Reduction to Number Partitioning}

The number partitioning problem is, given a list of positive integers $\delta_1,\ldots,\delta_n$ to find
\begin{equation}
\label{perfect}
\sum_{j=1}^n s_j \delta_j = 0
\end{equation}
where $s_j \in \{+1,-1\}$. The number partitioning problem is NP-complete\footnote{Technically, NP is a class of decision problems. The NP-complete version of the partitioning problem is to decide whether a solution to \eq{def} exists. However, by standard arguments \cite{Papadimitriou}, the decision and search versions of the problem are essentially equivalent; the complexity of finding a solution exceeds the complexity of deciding whether one exists by at most a factor of $n$.} and in fact was a member of the list of 21 problems shown to be NP-complete in the 1972 paper of Karp \cite{Karp72}, which together with Cook's 1971 paper \cite{Cook71} is credited with founding the theory of NP-completeness.

The problem of finding vacua in the ADK model with cosmological constant $10^{-122}$ differs superficially from the number partitioning problem in its standard form, but can easily be converted. To do so, first note that we can choose our labels so that for each $j$, $E_1^{(j)} \geq E_0^{(j)}$. Then, for each $j =1,\ldots n$ let
\begin{eqnarray}
\delta_j & = & (E_1^{(j)}-E_0^{(j)})/2 \label{deltaj} \\
\mu_j & = & (E_1^{(j)}+E_0^{(j)})/2.
\end{eqnarray}
In this notation, \eq{def} becomes
\begin{equation}
\label{lameq}
\Lambda = \delta_0 + \sum_{j=1}^n s_j \delta_j
\end{equation}
where
\begin{equation}
\delta_0 = \sum_{j=1}^n \mu_j. \label{delta0}
\end{equation}
It is clear that finding a solution to \eq{lameq} is very closely related to the number partitioning problem. There are three technical differences. First, the numbers involved are reals rather than integers. This is inconsequential, as reals can be scaled up and rounded to integers, with the scale factor determined by the needed level of precision. Henceforth, we will refer to both the problem of obtaining residue $\Lambda$ starting with real inputs of order 1 and the problem of obtaining residue 1 starting with integers of order $\Lambda^{-1}$ as number partitioning, as will be clear from context.

A second difference is that in many works on integer partitioning, one wishes to find a partition in which the residue is zero, rather than merely small. Third, in the problem arising from the ADK model, there is no variable $s_0 \in \{-1,+1\}$ multiplying $\delta_0$. Nevertheless, algorithms that were designed for solving the standard number partitioning problem can be easily adapted to this slight variant of the problem, as we now illustrate.

\subsection{Cost of Brute Force Search}
\label{cost}

Consider the number partitioning problem on real numbers, where problem instances are generated by drawing $n$ numbers independently at random from the uniform distribution on $[0,1]$. In \cite{KKLO86} it was proven that the median optimal residue is $\Theta( \sqrt{n} 2^{-n} )$. (The big-$\Theta$ notation indicates that the asymptotic scaling as $n \to \infty$ is $\sqrt{n} 2^{-n}$ up to constant factors.) Thus, for a solution with residue $\Lambda$ to exist, one needs $\sqrt{n} 2^{-n} \lesssim \Lambda$. One can show that asymptotically, this means the minimum viable value of $n$ scales as
\begin{equation}
\label{nsim}
n \sim \log_2 \Lambda^{-1} + \frac{1}{2} \log_2 \log_2 \Lambda^{-1}.
\end{equation}
To find a residue of size $\Lambda$ one needs to perform all arithmetic with at least
\begin{equation}
\label{bsim}
b \sim \log_2 \Lambda^{-1}
\end{equation}
bits of precision.

A naive method for brute force search would be to increment through all $2^n$ possible choices of sign $s_1,\ldots,s_n \in \{+1,-1\}$ and for each one, compute the corresponding sum, and compare it against the threshold for sufficient smallness (\emph{e.g.} $10^{-122}$). Such an algorithm would perform $n 2^n$ addition (or subtraction) operations, each on $b$ bits. Addition or subtraction of a pair of $b$-bit numbers can be done by a quantum circuit of $O(b)$ elementary gates \cite{VBE96, Draper, DKRS04, CDKM, Takahashi, Wang16}. Thus the total complexity of this algorithm is $O(nb 2^n)$.

However, there is a somewhat more efficient algorithm that still arguably qualifies as brute force search. Rather than summing up the residue from scratch with each new choice of signs, one could use the residue from the previous calculation and add or subtract $2\delta_j$ for each $j$ in which the sign has changed. For any $n$ there always exists an ordering of the $2^n$ bit strings of length $n$ such that each bit string is obtained from the previous one by only flipping a single bit. These orderings are called Gray codes, and they can furthermore be generated by efficient classical algorithms \cite{Doran}. By ordering the choices of sign according to a Gray code one thus has to do $n$ additions on the first step, and only one addition or subtraction on each of the subsequent $2^n -1$ steps. This brings the total complexity of the algorithm down to $O(b 2^n)$ elementary quantum gates. By \eq{nsim} and \eq{bsim} this yields a total complexity of order $\Lambda^{-1} \left(\log_2 \Lambda^{-1} \right)^{3/2}$.

\section{Algorithms for number partitioning}
\label{sec:alg}

In this section, we discuss efficient algorithms for the number partitioning problem.

The number partitioning problem is NP-complete. Assuming $P \neq NP$ this implies that no polynomial-time classical algorithm can solve all instances of number partitioning in time scaling polynomially in $n$. However, this does not forbid the existence of parameter regimes in which classical algorithms can solve the problem in polynomial time. In fact, for many NP-complete problems, including the canonical example of 3-SAT, randomly generated instances are efficiently solvable generically; exponentially hard instances require fine-tuning \cite{Monasson}.

Random instances of number partitioning have been well studied using methods of statistical mechanics. The standard ensemble of instances most typically studied is to set some magnitude parameter $B$ and then choose $n$ integers $\delta_1,\ldots,\delta_n$ independently uniformly at random from the range $\{1,2,\ldots, B\}$. If $\sum_{j=1}^n \delta_j \equiv 1 \textrm{ mod } 2$ then any sum of the form $\sum_{j=1}^n \pm \delta_j$ will be odd, and it is impossible for a solution to \eq{perfect} to exist. Thus, it is conventional to define a perfect partition as a solution to \eq{perfect} in the case that $\sum_{j=1}^n \delta_j$ is even, and as a solution to $\sum_{j=1}^n s_j \delta_j = 1$ in the case that $\sum_{j=1}^n \delta_j$ is odd. Whether a perfect partition exists for an instance of number partitioning sampled from the standard ensemble depends on the relationship between $n$ and $B$. If $n$ is too small relative to $B$ then the system is overconstrained and is likely to have no perfect partitions, whereas if $n$ is sufficiently large relative to $B$ then the system is underconstrained and is likely to have many perfect partitions. More precisely, as shown in \cite{Mertens_PT}, in the limit of large $n$, randomly generated number partitioning problems will have no perfect partitions for $B > 2^{n + O(\log n)}$ and will have exponentially many partitions for $B < 2^{n + O(\log n)}$. As is the case for many NP-complete problems, the number partitioning problem becomes easier for instances sufficiently far from the phase transition. 

For example, the Karmarkar-Karp algorithm solves number partitioning in time $O(n \log n)$ for $B < n^{c \log n}$, which is to say when $n > \exp \left[ \sqrt{\frac{\log B}{c} } \right]$ for some constant $c$. It was proven rigorously in \cite{Yakir} that $c = \frac{1}{2 \log 2} = 0.721\ldots$ suffices. In \sect{sec:experiments} we empirically achieve success with $c = 0.662$, which is in rough agreement with the empirical testing in \cite{BM08}. Nonetheless, the statistical mechanics arguments in \cite{BM08} suggest that $c = 0.721$ is the true asymptotic value as $n \to \infty$.

\subsection{The Karmarkar-Karp Algorithm}

\begin{figure}[htb]
\begin{center}
\includegraphics[width=0.7\textwidth]{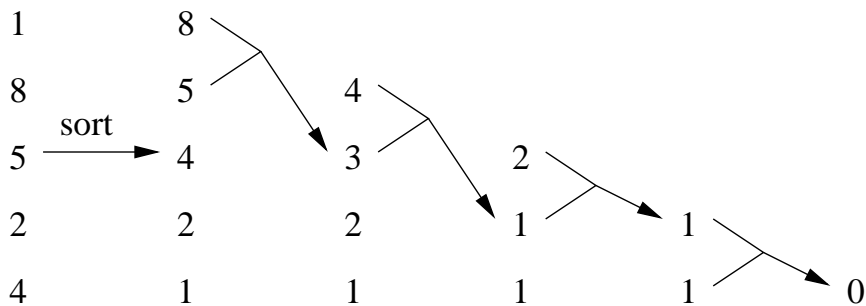}
\caption{\label{KKexamp}An example of the Karmarkar-Karp algorithm. At the first step the numbers are sorted. At each subsequent step, the largest two numbers are replaced by their difference, which is then inserted into the appropriate location in the list so that it remains sorted. The sequence of moves in the example shown finds the solution $1-(2-(4-(8-5)))=0$.}
\end{center}
\end{figure}

The Karmarkar-Karp algorithm is based on the intuition that the largest numbers should be given opposite sign in order to achieve cancellation. The Karmarkar-Karp strategy is to commit to giving the largest two numbers opposite signs without specifying which should be positive and which should be negative. This reduces the problem to a new instance of integer partitioning with one fewer number: the largest two numbers have been replaced by their difference. This is then treated in the same manner, until only one number is left, which is the final residue $\sum_{i=1}^n s_i \delta_i$. An example is given in figure \ref{KKexamp}. 

The initial sorting step has complexity $O(n \log n)$ by standard algorithms. Inserting a number into the correct location in an ordered list can be achieved with complexity $O(\log n)$ using a standard data structure called a heap \cite{Cormen}. There are exactly $n-1$ differencing-and-insertion steps needed to arrive at a final residue. Thus the total complexity of the algorithm is $O(n \log n)$.

The Karmarkar-Karp algorithm is heuristic in the sense that for some problem instances for which a perfect partition exists, the Karmarkar-Karp algorithm will fail to find it. On the other hand, as mentioned earlier, for random instances of integer partitioning with $B < n^{0.721 \log n}$, the Karmarkar-Karp algorithm will succeed with probability going to 1 as $n \to \infty$ \cite{Yakir}. Korf \cite{Korf} has introduced an extension of the Karmarkar-Karp algorithm, which initially proceeds identically to the Karmarkar-Karp algorithm and terminates if this yields a perfect partition. However, if it fails to find a perfect partition it continues searching by backtracking and trying assignments in which the largest two numbers are given the same sign. The details of Korf's algorithm are such that it is guaranteed to find a perfect partition provided one exists. For $B < n^{c \log n}$ Korf's algorithm matches the performance of the Karmarkar-Karp algorithm, but for $B \gg n^{c \log n}$ it may have exponentially long runtime.

Other heuristic algorithms derived from Karmarkar-Karp were studied in \cite{Ruml}, where it was empirically found that, in the regime where Karmarkar-Karp finds a residue much larger than the optimal residue, modest improvement in residue size can be obtained by exhaustively or stochastically searching solutions "nearby" to the Karmarkar-Karp solution, if the notion of nearness is carefully chosen. However, other than near the Karmarkar-Karp solution, the optimization landscape in number partitioning problems was found to be hard to distinguish from random, based on any of the neighborhood notions that were investigated. Thus there appears to be little structure in the problem for general-purpose optimization heuristics such as simulated annealing or genetic algorithms to exploit. This is corroborated by the relatively modest performance improvements obtained by such heuristics on number partitioning in other studies \cite{Johnson, Arguello, Berretta}.

In analyzing the performance of the Karmarkar-Karp algorithm it is standard to consider the ensemble of instances where the $\delta_1,\ldots,\delta_n$ are independent, identically distributed random variables, typically sampled from a uniform distribution on some range $0$ to $B$. The instances of number partitioning arising in the context of the ADK model may slightly differ from this. In particular, from equations \eq{deltaj} through \eq{delta0}, one sees that if $E_1,\ldots,E_n$ are each of order $B$, then $\delta_1,\ldots,\delta_n$ will be of order $B$, but $\delta_0$ will generically be of order $\sqrt{n}B$. It is easy to see that this makes only a small difference to the performance of the Karmarkar-Karp algorithm. The first $\sim \sqrt{n}$ differencing steps will all be used to difference from $\delta_0$. After that, one is left with a standard instance of integer partitioning in which all the numbers are of similar magnitude, and the Karmarkar-Karp algorithm performs as it would on the standard ensemble. Thus, whereas for the standard ensemble, one would have required a minimum of $n_{\min}^{\textrm{std}} \simeq \exp \left[ \sqrt{ \frac{\log \Lambda^{-1}}{c}} \right]$, the minimum number of fields in the ADK case may be slightly larger: $n_{\min}^{\textrm{ADK}} \simeq n_{\min}^{\textrm{std}} + \sqrt{n_{\min}^{\textrm{std}}}$.

In \sect{sec:experiments} we give the results of some computer experiments on the performance of the Karmarkar-Karp algorithm confirming the predictions of the statistical analyses referenced above, and giving a quantitative sense of the practical performance of the algorithm. For simplicity, and to facilitate comparison with the existing literature, the experiments in \sect{sec:experiments} are performed using a standard ensemble of instances of number partitioning.

\subsection{Dynamic programming}

The computational difficulty of the number partitioning problem depends on the number of numbers $n$, and their magnitudes. In the regime where the $B = \max_j \delta_j$ is only polynomially large, \emph{i.e.} the number of bits needed to represent the numbers scales only as some power of $\log n$, the number partitioning problem can be solved in polynomial time on classical computers using a standard technique called dynamic programming. Specifically, as is described nicely in \S 4.2 of \cite{GareyJohnson}, dynamic programming solves the number partitioning problem in time $\widetilde{O}(n D)$ where $D = \sum_{j=1}^n \delta_j$. Problems such as number partitioning that can be solved in polynomial time when all the input numbers are restricted to polynomial magnitude (rather than allowing them to be polynomially many bits long) are said to be pseudo-polynomial \cite{Mertens}.

\subsection{Adapting algorithms for subset sum}

Number partitioning, subset sum, and knapsack problems are all variants of essentially the same problem. Algorithms for one are often applicable, with minor modification, to the others. For example a straightforward meet-in-the-middle tree search \cite{horowitz1974computing} applies to all these problems and succeeds in finding the optimal residue in time $\approx 2^{0.5n}$. At present, the \emph{asymptotically} best upper bound on the classical complexity of finding the optimal solution to number partitioning problems is given by the algorithm of \cite{BCJ}, which is guaranteed to succeed in time $O(2^{0.291n})$. The asymptotically best upper bound on the quantum complexity of this problem given by the quantum algorithm of \cite{BJLM}, which is guaranteed to find the optimum using a number of elementary steps (quantum gates) at most $O(2^{0.241n})$. (This quantum algorithm is based on quantum walks. An adiabatic quantum algorithm for this problem has also been analyzed, but its runtime is not known. Numerical calculations in \cite{Denchev} suggest a runtime scaling as $2^{0.8n}$. The adiabatic algorithm may also be limited in its capacity to accommodate large $B$.)

As discussed in section \ref{cost}, the minimum value of $n$ such that the number partitioning problem is likely to have a solution of order $\Lambda$ is asymptotically $\log_2 \Lambda^{-1} + \frac{1}{2} \log_2 \log_2 \Lambda^{-1}$. The algorithm of \cite{BCJ} could solve a problem of this size with runtime of order $\left(\Lambda^{-1} \right)^{0.291} \left( \log_2 \Lambda^{-1} \right)^{0.146}$. 

\subsection{Adapting lattice sieves}

\begin{figure}[b]
\begin{tikzpicture}
\begin{axis}[
  xlabel= $b$ (Block size),
  ylabel= $\log_2 s$ (Residue size),
  ylabel near ticks,
  legend pos = north east,
  legend style = {font=\small, no markers}]
\addplot+[domain=10:48]{0.37*ln(x)/ln(2) - x + 2.33};
\addlegendentry{$s = 5.0b^{0.37}2^{-b}$}
\addplot table [x=N, y=S]{output.log.exponential};
\addlegendentry{exponential input}
\addplot table [x=N, y=S]{output.log.uniform};
\addlegendentry{uniform input}
\end{axis}
\end{tikzpicture}
\caption{Expected relative optimal residue size versus input size for number partitioning problems on a block of $b$ random numbers with the specified distribution. For each $b$ in this given range, mean size for 1000 experiments is shown. Each experiment generated high precision floating point input data with mean one and a complete NPP solver produced the optimal residue. Assuming the distribution of optimal residues is exponential, the maximum likelihood estimator of the mean is biased, hence the least square error estimator was used to find the mean in each case. The model parameters in $s = 5.0b^{0.37}2^{-b}$ were generated by linear regression on the data with uniformly distributed inputs.}
\label{fig:residues}
\end{figure}
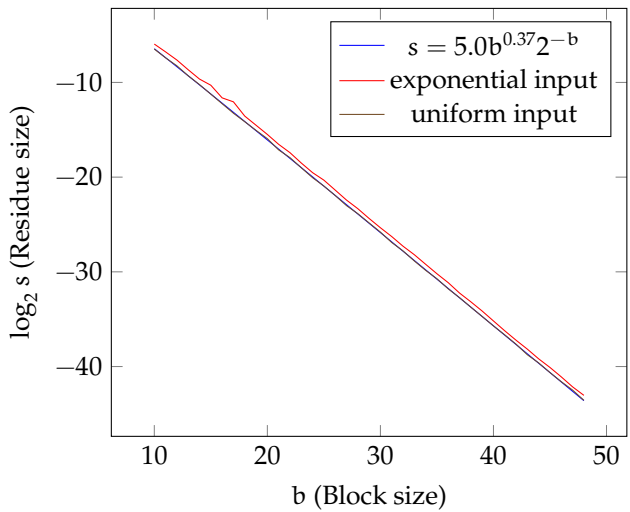

Here we explore a very simple sieve mechanism for solving the number partitioning problem inspired by ``lattice sieves'' \cite{ajtai2001sieve}. The Karmarkar-Karp algorithm can be viewed as a form of the Gauss sieve \cite{micciancio2010faster} for a 1-dimensional lattice. Curiously, while more sophisticated lattice sieves easily outperform the Gauss sieve on high dimensional lattices \cite{zhang2013three, becker2014sieve, laarhoven2015faster, becker2015speeding, bai2016tuple}, here we find this is seemingly not the case for the number partitioning problem. The simple sieve we present here is similar in spirit to the ``tuple sieve'' of \cite{bai2016tuple}, but cannot match the performance of the Karmarkar-Karp algorithm as we will show.  Nonetheless, the key advantage of this style of sieve is that it is not restricted to the number partitioning problem and so could be easily adapted to other models of the landscape.

\begin{figure}
\begin{tabular}{c@{}c}
\begin{tikzpicture}
\begin{axis}[
  scale = 0.8,
  title = {$b = 10$},
  title style = {at={(0.3,0.45)}},
  xlabel=$\log_2$ Size,
  ylabel={\footnotesize Cumulative Likelihood},
  ylabel near ticks,
  legend pos = north west,
  legend style = {font=\tiny, no markers}]

\addplot table [x=e, y=x, mark=none]{output.10.exponential-plot};
\addlegendentry{data (exponential)}

\addplot table [x=e, y=m, mark=none]{output.10.exponential-plot};
\addlegendentry{model (exponential)}

\addplot table [x=e, y=x, mark=none]{output.10.uniform-plot};
\addlegendentry{data (uniform)}

\addplot table [x=e, y=m, mark=none]{output.10.uniform-plot};
\addlegendentry{model (uniform)}

\end{axis}
\end{tikzpicture}
&
\begin{tikzpicture}
\begin{axis}[
  scale = 0.8,
  title = {$b = 20$},
  title style = {at={(0.3,0.45)}},
  xlabel=$\log_2$ Size,
  legend pos = north west,
  legend style = {font=\tiny, no markers}]

\addplot table [x=e, y=x, mark=none]{output.20.exponential-plot};
\addlegendentry{data (exponential)}

\addplot table [x=e, y=m, mark=none]{output.20.exponential-plot};
\addlegendentry{model (exponential)}

\addplot table [x=e, y=x, mark=none]{output.20.uniform-plot};
\addlegendentry{data (uniform)}

\addplot table [x=e, y=m, mark=none]{output.20.uniform-plot};
\addlegendentry{model (uniform)}

\end{axis}
\end{tikzpicture}
\\
\begin{tikzpicture}
\begin{axis}[
  scale = 0.8,
  title = {$b = 30$},
  title style = {at={(0.3,0.45)}},
  xlabel=$\log_2$ Size,
  ylabel={\footnotesize Cumulative Likelihood},
  ylabel near ticks,
  legend pos = north west,
  legend style = {font=\tiny, no markers}]

\addplot table [x=e, y=x, mark=none]{output.30.exponential-plot};
\addlegendentry{data (exponential)}

\addplot table [x=e, y=m, mark=none]{output.30.exponential-plot};
\addlegendentry{model (exponential)}

\addplot table [x=e, y=x, mark=none]{output.30.uniform-plot};
\addlegendentry{data (uniform)}

\addplot table [x=e, y=m, mark=none]{output.30.uniform-plot};
\addlegendentry{model (uniform)}

\end{axis}
\end{tikzpicture}
&
\begin{tikzpicture}
\begin{axis}[
  scale = 0.8,
  title = {$b = 40$},
  title style = {at={(0.3,0.45)}},
  xlabel=$\log_2$ Size,
  legend pos = north west,
  legend style = {font=\tiny, no markers}]

\addplot table [x=e, y=x, mark=none]{output.40.exponential-plot};
\addlegendentry{data (exponential)}

\addplot table [x=e, y=m, mark=none]{output.40.exponential-plot};
\addlegendentry{model (exponential)}

\addplot table [x=e, y=x, mark=none]{output.40.uniform-plot};
\addlegendentry{data (uniform)}

\addplot table [x=e, y=m, mark=none]{output.40.uniform-plot};
\addlegendentry{model (uniform)}

\end{axis}
\end{tikzpicture}
\end{tabular}
\caption{Plots of cumulative likelihood of observing the optimal residue versus (log) size of the optimal residue. The model is the cumulative distribution function of the exponential distribution where the single parameter $\lambda$ is computed from the data using the least squares estimator. For block sizes $b=10,20,30,40$, each plot was generated from high precision floating point input data (uniformly or exponentially distributed) with mean one and a complete solver produced the optimal residue.}
\label{fig:sieve-model-data}
\end{figure}
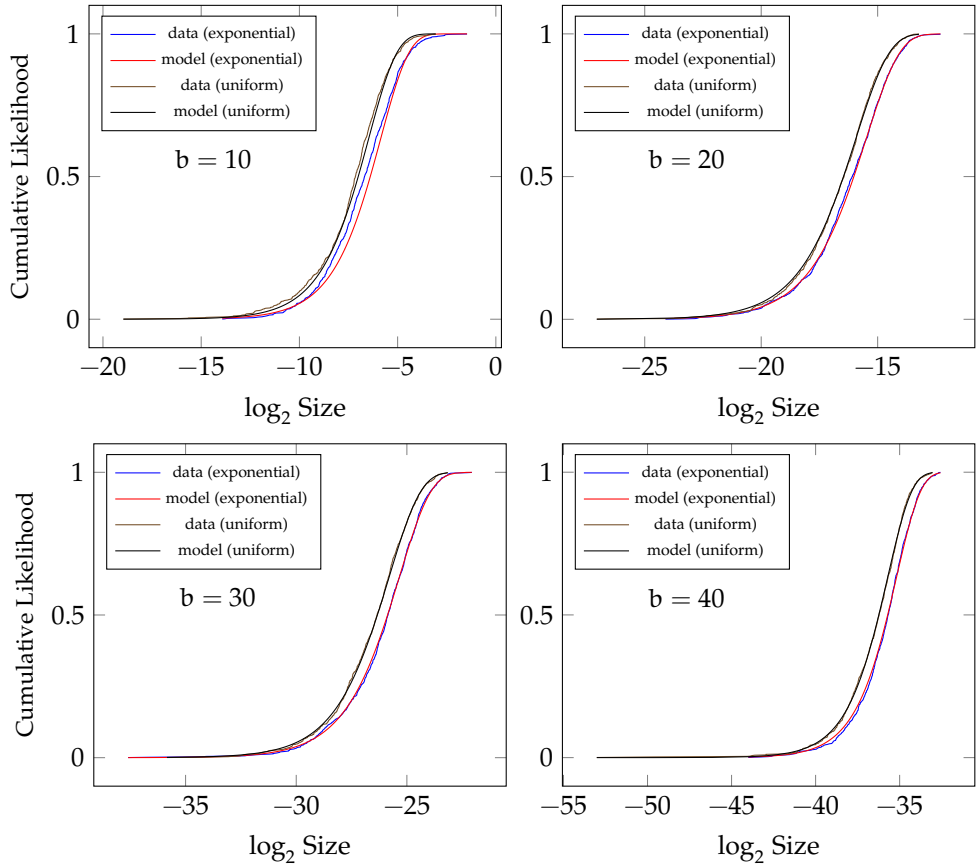

In general, a sieve consists of several stages. For us, the input to a stage is a collection of numbers; these are partitioned into small blocks of size $b$ and on each of these blocks the number partition problem is solved for the optimal residue. This collection of residues is the output of the sieve stage, which then becomes the input for the next stage. There are number of algorithms to solve for the optimal residue, some of which are illustrated in the previous sections. All of these take work $2^{\alpha b + o(b)}$. As long as the distribution of the input data is sufficiently well behaved, the optimal residues will be exponentially distributed with expected size $2^{-b + o(b)}$, asymptotically $O(\sqrt{b}2^{-b})$ \cite{KKLO86, mertens2000random, bauke2004number}. In figure \ref{fig:residues}, we validate this scaling for small $b$ but recover a smaller power in the polynomial factor in this formula. In figure \ref{fig:sieve-model-data}, we also validate that the distribution of the residues is well-modeled as exponential with the parameter $\lambda$ estimated from the data.

If our input is $n$ fields producing mean energy differences $\delta_j \approx 1$, the first sieve stage involves solving $n/b_1$ number partition problems, each of size $b_1$. The work for this stage is $\approx \frac{n}{b_1} 2^{\alpha b_1}$ and the output is $\frac{n}{b_1}$ residues exponentially distributed with mean size $\approx 2^{-b_1}$. The second sieve stage partitions these into blocks of size $b_2$ and solves solves the number partition problem on each to produce $\frac{n}{b_1b_2}$ residues of size $\approx 2^{-(b_1+b_2)}$. And so on.

The goal is that after $k$ sieve stages we produce a single residue of expected length $2^{-t} \approx 2^{-(b_1 + \cdots + b_k)}$. The optimal work is given when we follow an ``equipartition principle'' and balance the amount of work done on each sieve stage. For example, the first sieve stage involves solving many more number partition problems than the second stage, and so we should choose $b_2 > b_1$ so as to balance the amount of work done during the first stage with that done in the second. Specifically, in stage $j \leq k$ of the sieve, we solve $n/(b_1\cdots b_j)$ number partition problems with an overall work of $n/(b_1\cdots b_j)2^{\alpha b_j}$, which we balance with the work in stage $j-1$:
\begin{equation}
    \frac{n}{b_1\cdots b_j} 2^{\alpha b_j} \approx \frac{n}{b_1\cdots b_{j-1}} 2^{\alpha b_{j-1}}.
\end{equation}
Therefore we select $b_j$ implicitly by solving
\begin{equation}
    b_j - \tfrac{1}{\alpha}\log_2(b_j) \approx b_{j-1}.
\end{equation}
The overall work of the sieve is then $\sim \frac{kn}{b_1}2^{\alpha b_1}$. Examples of sieves for $k=2, \dots, 8$ stages, $\alpha = 0.5$, all targeting residues of size $\approx 2^{-400}$, is given in table \ref{table:nppsieve}.

This table indicates that the only sieves with $k=2,3,4$ can outperform Karmarkar-Karp in terms of the number of fields, which requires $n \approx 8\times 10^8$ to produce residues of size $\approx 2^{-400}$. At this size Karmarkar-Karp takes work roughly $2^{35}$, well below that of any of these sieves. To outperform Karmarkar-Karp with this style of sieve, the algorithm that solves number partitioning on the blocks would need to have $\alpha \lesssim 0.22$, and even then lower-order terms not counted in the asymptotic expression would likely dominate the work.

\begin{table}[t]
\begin{tabular}{ccccl}
$k$ & $n$ & $t$ & $w$ & $(n_1,n_2,\dots)$\\\hline
2 & $4.22\times 10^4$ & 400.0 & 107.62 & (198, 213)\\
3 & $2.65\times 10^6$ & 400.8 & 78.32 & (124, 139, 154) \\
4 & $1.19\times 10^8$ & 400.8 & 65.07 & (85, 98, 113, 126)\\
5 & $3.96\times 10^9$ & 400.0 & 58.14 & (59, 72, 85, 98, 112)\\
6 & $1.03\times 10^{11}$ & 400.3 & 54.53 & (41, 53, 65, 77, 91, 104) \\
7 & $1.97\times 10^{12}$ & 400.8 & 52.70 & (27, 38, 49, 61, 74, 87, 100) \\
8 & $2.54\times 10^{13}$ & 400.5 & 51.88 & (16, 26, 36, 48, 59, 72, 85, 98)\\
\end{tabular}
\caption{Example sieves for $k=2,\dots,8$ stages with overall expected residue $2^{-t} \approx 2^{-400}$. The size of the blocks $(n_1,n_2,\dots)$ are selected so the overall work in each stage is approximately equal. In this range as the number of layers increases, the required number of input fields $n$ increases, and the overall work of the sieve $2^w$ decreases. However at smaller block sizes (for instance $b_1 = 16$ for $k=8$), variations in the size of the resulting residues is large and so the work estimates given are less accurate.}\label{table:nppsieve}
\end{table}

\section{Computer Experiments}
\label{sec:experiments}

In this section, we apply fast algorithms to the problem of finding a small cosmological constant in an ADK landscape. We show that they allow the Computational Censorship Hypothesis to be satisfied.

\subsection{Karmarkar-Karp}

\begin{figure}
\begin{center}
\includegraphics[width=0.6\textwidth]{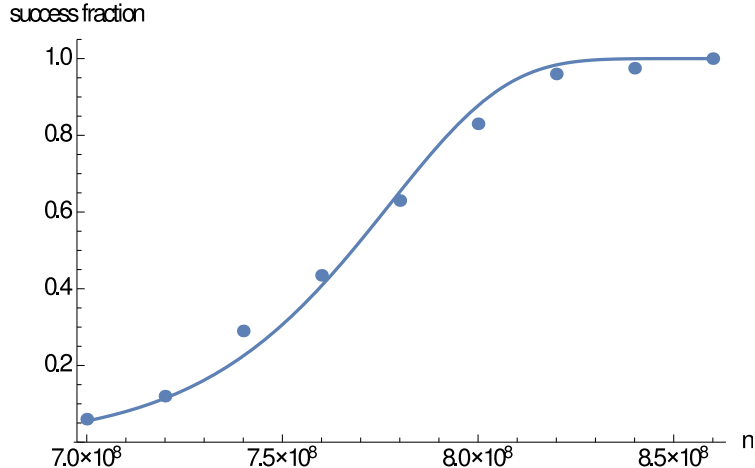}
\caption{At each value of $n$, 200 instances of number partitioning are generated with each of the $n$ numbers independently sampled uniformly from $\{0,1,2,\ldots,2^{430}-1\}$. The fraction of instances in which the Karmarkar-Karp algorithm found a residue smaller than $2^{30}$ is shown for each $n$. The theoretically predicted success probability of $1 - \exp \left [-\frac{e^{c \log(n)^2}}{2^{400}} \right]$ is also shown, with $c =0.6615$ determined by fitting to the data. The asymptotic value of $c$ as $n \to \infty$ is predicted to be $1/\sqrt{2} \simeq 0.7071$.\label{fig:fitplot}}
\end{center}
\end{figure}

To empirically test the Karmarkar-Karp algorithm in a regime relevant to the cosmological constant problem, we generated random instances of the number partitioning problem, at various values of $n$ in which each of the $n$ numbers are independently sampled uniformly from $\{0,1,2,\ldots,2^{430}-1\}$. In figure \ref{fig:fitplot}, we plot the fraction of instances on which the Karmarkar-Karp algorithm was successful with $n$ numbers, where we defined success as achieving residue less than $2^{30}$. In the context of finding small cosmological constant within the ADK model, one starts with real numbers of order 1, and seeks to find a residue of order $10^{-122}$. Here we have scaled up the numbers by a factor of $2^{430}$ and represented them as integers. This use of fixed-point arithmetic is strictly for computational convenience. Our definition of success corresponds to achieving a residue which is smaller than the magnitude of the initial numbers by a factor of $2^{400} \simeq 10^{120}$ and thus corresponds to finding a cosmological constant close to that observed for our universe\footnote{A more precise match to our universe would be to seek a factor of $2^{406}$, but this was not convenient to work with because it put the memory requirements of the algorithm just slightly beyond the available 128G of RAM on most of our computers. Achieving a factor $2^{406}$ requires $n \simeq 8.7 \times 10^8$ and correspondingly an increase in time and memory cost of less than $20\%$.}. The extra 30 bits of precision are to ensure that ``numerical noise'' should be small.

By the analysis of \cite{BM08}, if the Karmarkar-Karp algorithm is applied to real numbers uniformly distributed on $[0,1]$, the size of the final residue should be exponentially distributed. That is, the probability that the residue lies between $y$ and $y + dy$ should be $\lambda e^{-\lambda y}dy$, where
\begin{equation}
\lambda=e^{-c \log^2 n}
\end{equation} 
and $c$ asymptotically equal to $1/\sqrt{2}$ as $n \to \infty$. Empirical studies at finite $n$ consistently observe values of $c$ smaller than $1/\sqrt{2}$ \cite{BM08}. By defining success to be a reduction factor of $\epsilon = 2^{-400}$, we should obtain success probability
\begin{eqnarray}
P & = & \int_0^\epsilon \lambda e^{- \lambda y} dy \\
  & = & 1-\exp \left[ -e^{-c \log^2 n} \epsilon \right]. \label{eq:p}
\end{eqnarray}
As one can see from figure \ref{fig:fitplot}, the observed success fraction from our trials of the Karmarkar-Karp algorithm on random instances agrees well with this prediction if we take $c = 0.6615$.

\subsection{Sieves}

The predicted work of a sieve to produce a residue of length $2^{-400}$ is not so large that the universe would be unable to compute it, but it is large enough to require significant effort with current hardware. As a simple proof of concept, we will tackle a scaled down version with four sieve stages of block sizes $(b_1,b_2,b_3,b_4) = (20, 30, 40, 50)$, and use a simple meet-in-the-middle algorithm ($\alpha = 0.5$) to solve the number partitioning problem \cite{horowitz1974computing}. The profile of this experiment is as follows, which predicts an expected size of the final residue output at sieve stage four to be $\mathbb{E}[s] = 2^{-121.3}$.

\begin{center}
\begin{tabular}{|r|c|rl|c|c|c|}\hline
Stage  & $b$ & Inputs & Distribution & \#NPPs & Work & $\mathbb{E}[s]$\\\hline
One    & $20$ & 1200000 & Uniform & 60000 & $2^{25.9}$ & $2^{-16.1}$\\
Two    & $30$ & 60000 & Exponential & 2000 & $2^{26.0}$ & $2^{-41.3}$\\
Three  & $40$ & 2000 & Exponential & 50 & $2^{25.6}$ & $2^{-76.4}$\\
Four   & $50$ & 50 & Exponential & 1 & $2^{25.0}$ & $2^{-121.3}$\\\hline
\end{tabular}
\end{center}

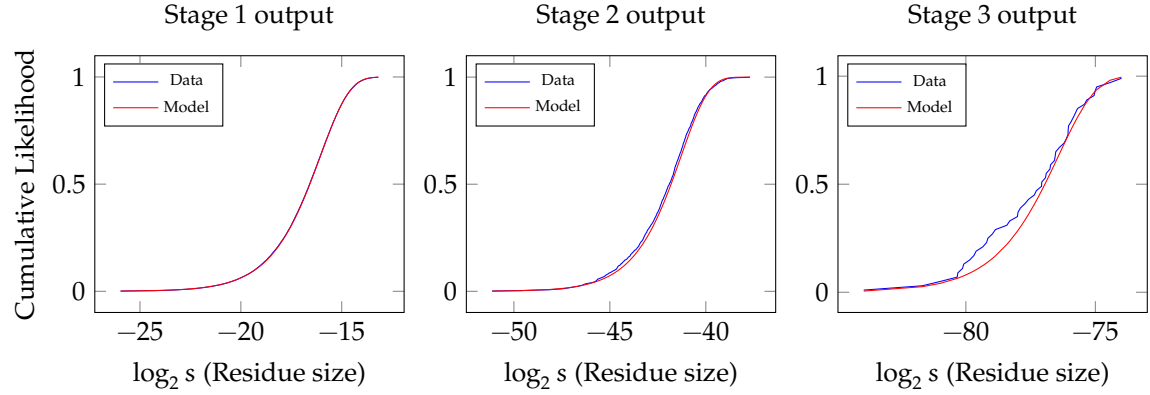
\begin{figure}
\begin{tabular}{ccc}
\begin{tikzpicture}
\begin{axis}[
  scale = 0.6,
  title = {Stage 1 output},
  title style = {at={(0.5,1)}},
  xlabel={\small $\log_2 s$ (Residue size)},
  ylabel={\small Cumulative Likelihood},
  ylabel near ticks,
  legend pos = north west,
  legend style = {font=\tiny, no markers}]

\addplot table [x=e, y=x, mark=none]{layer.1-plot};
\addlegendentry{Data}
\addplot table [x=e, y=m, mark=none]{layer.1-plot};
\addlegendentry{Model}
\end{axis}
\end{tikzpicture}
&
\begin{tikzpicture}
\begin{axis}[
  scale = 0.6,
  title = {Stage 2 output},
  title style = {at={(0.5,1)}},
  xlabel={\small $\log_2 s$ (Residue size)},
  ylabel near ticks,
  legend pos = north west,
  legend style = {font=\tiny, no markers}]

\addplot table [x=e, y=x, mark=none]{layer.2-plot};
\addlegendentry{Data}
\addplot table [x=e, y=m, mark=none]{layer.2-plot};
\addlegendentry{Model}
\end{axis}
\end{tikzpicture}
&
\begin{tikzpicture}
\begin{axis}[
  scale = 0.6,
  title = {Stage 3 output},
  title style = {at={(0.5,1)}},
  xlabel={\small $\log_2 s$ (Residue size)},
  ylabel near ticks,
  legend pos = north west,
  legend style = {font=\tiny, no markers}]

\addplot table [x=e, y=x, mark=none]{layer.3-plot};
\addlegendentry{Data}
\addplot table [x=e, y=m, mark=none]{layer.3-plot};
\addlegendentry{Model}
\end{axis}
\end{tikzpicture}
\end{tabular}
\caption{Plots of cumulative likelihood of observing the optimal residue versus (log) size of the optimal residue for a four stage sieve. Input to stage one was $n = 1.20\times 10^6$ mean one uniformly distributed numbers. Stage one combined $b_1=20$ numbers in each number partitioning problem to produce $60000$ optimal residues, forming the input to stage two. Stage two combined $b_2=30$ numbers in each problem to produce $2000$ optimal residues. Stage three combined $b_3=40$ numbers to produce $50$ optimal residues. Finally stage four combined these $b_4 = 50$ numbers to produce an overall residue of $6.54\times 10^{-38}$. This final residue was slightly smaller than the predicted $2^{-121.3}$. The sieve completed in $152$ seconds on a standard desktop computer.}
\label{fig:simple-sieve}
\end{figure}

The result of the experiment is captured in figure \ref{fig:simple-sieve}.

\section*{Acknowledgements}
We would like to thank Scott Aaronson, Adam Bouland, and Liam McAllister for discussions. N.B.\ is supported in part by the DuBridge Fellowship of the Walter Burke Institute for Theoretical Physics. R.B.\ is supported in part by the Berkeley Center for Theoretical Physics, by the National Science Foundation (award numbers PHY-1521446, PHY-1316783), by FQXi, and by the US Department of Energy under contract DE-AC02-05CH11231. S.J.\ and B.L.\ thank U. Maryland for use of the {\em Deepthought2} high performance computing cluster. Parts of this manuscript are a contribution of NIST, an agency of the US government, and are not subject to US copyright.

\clearpage

\bibliography{landscape,all}

\end{document}